\documentclass[12pt]{iopart}
\usepackage{iopams}
\usepackage{graphicx}
\input amssym.def \input amssym

\begin{document}

\title[]{Five-Qubit Contextuality, Noise-Like Distribution of Distances Between Maximal Bases and Finite Geometry}

\author{ Michel Planat$^{1}$
and Metod Saniga$^{2}$
}

\address{$^1$Institut FEMTO-ST, CNRS, 32 Avenue de
l'Observatoire,\\ F-25044 Besan\c con, France }
\ead{michel.planat@femto-st.fr}
\vspace*{.1cm}

\address{$^2$Astronomical Institute, Slovak Academy of Sciences,\\ SK-05960 Tatransk\'{a} Lomnica, Slovak Republic}
\ead{msaniga@astro.sk}
\vspace*{.1cm}

\begin{abstract}
Employing five commuting sets of five-qubit observables, we propose specific $160-661$ and $160-21$ state proofs of the Bell-Kochen-Specker theorem that are also proofs of Bell's theorem.
 A histogram of the `Hilbert-Schmidt' distances between the corresponding maximal bases shows in both cases a noise-like behaviour. The five commuting sets are also ascribed a finite-geometrical meaning in terms of the structure of symplectic polar space $W(9,2)$.

\end{abstract}


\noindent

\section{Introduction}

Projector-valued (also called von-Neumann) measurements on a $d$-dimensional quantum system are contextual if there is no way of assigning definite outcomes to yes-no tests that might be performed on a set of $n$ mutually compatible projection operators. Compatible measurements may have been realized in the past or may be realized in the future; that is, quantum contextuality is counterfactual by flouting the causality constraint \cite{Peres}. Contextuality also encompasses non-locality which requires not only  compatible, but also space-like separated tests.

A landmark statement for contextuality is the Bell-Kochen-Specker (BKS) theorem \cite{Bell1966,Kochen1967}, which may be formulated as follows. In a Hilbert space of dimension $d \ge 3$, it is always possible to find a finite set of rays/vectors that cannot each be assigned the value $1$ (for true) or $0$ (for false) such that, first, no two orthogonal rays are both assigned the value $1$ and, second, in any complete basis not all the rays are assigned the value $0$ \cite{Aravind1999}; these constraints are sometimes referred to as a non-coloring property of a BKS set.
The BKS theorem is closely related to Bell's theorem which is the statement that local realistic theories are in conflict with quantum mechanics. It was found that both theorems, viz. the BKS theorem about contextuality and Bell's theorem about non-locality, can be given a simultaneous proof provided that the selected uncolorable set of rays is complete (in the sense made explicit in \cite{Aravind1999}). For multiple qubits, the completeness argument means that each ray (usually represented by a column vector) is paired with a partner ray (obtained by inverting the column and flipping the signs).

Many proofs of the BKS theorem rely on magic geometrical configurations involving only operators and parity rules \cite{Mermin1993}. Along this line of action, the well-known Mermin square (for two qubits, $d=4$) and Mermin pentagram (for three qubits, $d=8$) were among the first to serve as an operator proof of the BKS theorem and Bell's theorem as well. This stems from the fact that the $24=6 \times 4$ rays originating from the $6$ commuting sets in the Mermin square, as well as  the $40=5 \times 8$ operators originating from the $5$ commuting sets of the Mermin pentagram, contain both a ray and its partner, having thus the required completeness. In the same vein of research, a four-qubit magic rectangle found by Harvey and Chryssanthacopoulos \cite{Harvey2008,SanPla1} may be used for both an operator and a state proof of the BKS theorem, as well as for a proof of Bell's theorem. Moreover, the found magic rectangle is similar to the pentagram \cite[(15)]{Planat2011}, with four operator bases of size five and one of size four; there are $80$ real states shared by these bases and one can find a non-parity BKS proof with only $21$ selected maximal bases.

In this paper, we extend a recent series of small proofs of the BKS theorem  \cite{Planat2011}, which were based on real rays/vectors associated with specific sets of two-, three- and four-qubit operators within the corresponding generalized Pauli group, to a five-qubit system. The magic configuration that motivated our study is a particular one from a sequence proposed by Aravind for odd Hilbert space dimensions \cite{Aravind2002}; see also \cite{DiVPeres}. Among the novelties we find (a) a non-parity proof of the BKS theorem with $160$ rays on $21$ maximal bases, (b) a noise-like distribution of the `Hilbert-Schmidt' distances between the bases and (c) a remarkable finite geometry underlaid by a hyperbolic quadric $Q^{+}(9,2)$ of the symplectic polar space $W(9,2)$.

\section{Five-qubit proofs of the BKS theorem}
\label{results}

\subsection{The magic five-qubit configuration and associated state BKS proofs}
The magic configuration we start with is a modification of that proposed by Aravind \cite{Aravind2002} and DiVincenzo and Peres \cite{DiVPeres}, namely:
\begin{eqnarray}
\{A_1,A_2,A_3,A_4,A_5;Z^5\},~~\{Z_1,Z_2,Z_3,Z_4,Z_5;Z^5\},\nonumber \\
\{A_1,A_3,Z_2,Z_4,X_5;X_1\},~~\{A_2,A_4,Z_3,Z_5,X_2;X_1\}, \nonumber \\
\{A_5,Z_1,X_2;X_5\}.
\label{BKSoperator}
\end{eqnarray}
Here, $Z_1 \equiv Z \otimes I \otimes I \otimes I \otimes I$, $Z_2 \equiv I \otimes Z \otimes I \otimes I \otimes I$,\ldots, $Z_5 \equiv I \otimes I \otimes I \otimes I \otimes Z$, $Z^5 \equiv
Z \otimes Z \otimes Z \otimes Z \otimes Z$, similarly for $X$, $A_1 \equiv X \otimes Z \otimes X \otimes I \otimes I$,  $A_2 \equiv I \otimes X \otimes Z \otimes X \otimes I$,\ldots, $A_5 \equiv
Z \otimes X \otimes I \otimes I \otimes X$, and
\begin{eqnarray*}
I = \left(
\begin{array}{cc}
1 & 0 \\
0 & 1 \\
\end{array}
\right),~
X = \left(
\begin{array}{cc}
0 & 1 \\
1 & 0 \\
\end{array}
\right),~
Y = \left(
\begin{array}{rr}
0 & -1 \\
1 & 0 \\
\end{array}
\right)
~{\rm and}~
Z = \left(
\begin{array}{rr}
1 & 0 \\
0 & -1 \\
\end{array}
\right).
\end{eqnarray*}
The operators preceding the semicolon in each set of pairwise commuting operators of (\ref{BKSoperator}) multiply to the last one {\it except for} the first set where the first five operators multiply to $-Z^5$. There are altogether $14$ operators and each of them occurs in exactly two commuting sets. Since each operator has the eigenvalues $\pm 1$, there is no way of assigning multiplicative properties to all  the eigenvalues while keeping the same multiplicative properties for the operators; hence, these five sets furnish an operator/observable proof of the BKS theorem.

There are $4 \times 32+8=136$ eigenstates associated with the five operator bases in (\ref{BKSoperator}). The states/rays in question form $85$ maximal bases that exhibit a bicoloring and, hence, do not lead to a state proof of the BKS theorem.
For such a proof, we have to pass to a slightly different set (which, however, is no longer ``magic"). In particular, in each of the first four sets of (\ref{BKSoperator})
we drop the observable following the semicolon and in the last set we replace the last observable by a couple of new ones, namely $Z_4$ and $A_3$:
\begin{eqnarray}
&\{A_1,A_2,A_3,A_4,A_5\} \equiv A,~~\{Z_1,Z_2,Z_3,Z_4,Z_5\} \equiv B,\nonumber \\
&\{A_1,A_3,Z_2,Z_4,X_5\} \equiv A',~~\{A_2,A_4,Z_3,Z_5,X_2\} \equiv B', \nonumber \\
&\{A_5,Z_1,X_2,Z_4,A_3\} \equiv C.
\label{BKSstate}
\end{eqnarray}
The $5\times 32=160$ eigenstates/rays form $661$ maximal bases that do not have a bicoloring. The simplest BKS proof we found contains, however, only $21$ maximal bases. We looked at the graph whose vertices are theses 21 bases and edges join two bases whenever they overlap (in one or several rays) and found that its automorphism group is isomorphic to $\mbox{aut}=\mathbb{Z}_2^5 \rtimes \mathbb{Z}_6$. The $160$ rays shared by the five commuting sets in (\ref{BKSstate}) explicitly read:

\scriptsize
\begin{eqnarray*}
&1:[1 0 0 0 0 0 0 0 0 0 0 0 0 0 0 0 0 0 0 0 0 0 0 0 0 0 0 0 0 0 0 0],
2:[0 1 0 0 0 0 0 0 0 0 0 0 0 0 0 0 0 0 0 0 0 0 0 0 0 0 0 0 0 0 0 0],\nonumber \\
&\cdots,32:[0 0 0 0 0 0 0 0 0 0 0 0 0 0 0 0 0 0 0 0 0 0 0 0 0 0 0 0 0 0 0 1],\nonumber \\
&33:[ 1  0  0  1  0 \bar{1}  1  0  0  1  1  0 \bar{1}  0  0  1  0 \bar{1}  1  0  1  0  0  1  1  0  0 \bar{1}  0 \bar{1} \bar{1}  0],
34:[ 1  0  0 \bar{1}  0 \bar{1} \bar{1}  0  0 \bar{1}  1  0  1  0  0  1  0 \bar{1} \bar{1}  0  1  0  0 \bar{1} \bar{1}  0  0 \bar{1}  0  1 \bar{1}  0],\nonumber \\
&35:[ 0  1  1  0 \bar{1}  0  0  1  1  0  0  1  0 \bar{1}  1  0  1  0  0 \bar{1}  0 \bar{1} \bar{1}  0  0 \bar{1}  1  0  1  0  0  1],
36:[ 0  1  1  0  1  0  0 \bar{1}  1  0  0  1  0  1 \bar{1}  0  1  0  0 \bar{1}  0  1  1  0  0 \bar{1}  1  0 \bar{1}  0  0 \bar{1}],\nonumber \\
&37:[ 1  0  0 \bar{1}  0  1  1  0  0  1 \bar{1}  0  1  0  0  1  0 \bar{1} \bar{1}  0 \bar{1}  0  0  1  1  0  0  1  0  1 \bar{1}  0],
38:[ 0  1  1  0 \bar{1}  0  0  1 \bar{1}  0  0 \bar{1}  0  1 \bar{1}  0 \bar{1}  0  0  1  0  1  1  0  0 \bar{1}  1  0  1  0  0  1],\nonumber \\
&39:[ 0  1 \bar{1}  0 \bar{1}  0  0 \bar{1}  1  0  0 \bar{1}  0 \bar{1} \bar{1}  0  1  0  0  1  0 \bar{1}  1  0  0 \bar{1} \bar{1}  0  1  0  0 \bar{1}],
40:[ 1  0  0 \bar{1}  0  1  1  0  0 \bar{1}  1  0 \bar{1}  0  0 \bar{1}  0 \bar{1} \bar{1}  0 \bar{1}  0  0  1 \bar{1}  0  0 \bar{1}  0 \bar{1}  1  0],\nonumber \\
&41:[ 1  0  0 \bar{1}  0 \bar{1} \bar{1}  0  0 \bar{1}  1  0  1  0  0  1  0  1  1  0 \bar{1}  0  0  1  1  0  0  1  0 \bar{1}  1  0],
42:[ 1  0  0  1  0 \bar{1}  1  0  0  1  1  0 \bar{1}  0  0  1  0  1 \bar{1}  0 \bar{1}  0  0 \bar{1} \bar{1}  0  0  1  0  1  1  0],\nonumber \\
&43:[ 1  0  0 \bar{1}  0  1  1  0  0  1 \bar{1}  0  1  0  0  1  0  1  1  0  1  0  0 \bar{1} \bar{1}  0  0 \bar{1}  0 \bar{1}  1  0],
44:[ 0  1  1  0 \bar{1}  0  0  1  1  0  0  1  0 \bar{1}  1  0 \bar{1}  0  0  1  0  1  1  0  0  1 \bar{1}  0 \bar{1}  0  0 \bar{1}],\nonumber \\
&45:[ 1  0  0  1  0  1 \bar{1}  0  0  1  1  0  1  0  0 \bar{1}  0  1 \bar{1}  0  1  0  0  1 \bar{1}  0  0  1  0 \bar{1} \bar{1}  0],
46:[ 0  1 \bar{1}  0  1  0  0  1 \bar{1}  0  0  1  0 \bar{1} \bar{1}  0 \bar{1}  0  0 \bar{1}  0 \bar{1}  1  0  0 \bar{1} \bar{1}  0 \bar{1}  0  0  1],\nonumber \\
&47:[ 1  0  0  1  0 \bar{1}  1  0  0 \bar{1} \bar{1}  0  1  0  0 \bar{1}  0 \bar{1}  1  0  1  0  0  1 \bar{1}  0  0  1  0  1  1  0],
48:[ 1  0  0 \bar{1}  0 \bar{1} \bar{1}  0  0  1 \bar{1}  0 \bar{1}  0  0 \bar{1}  0  1  1  0 \bar{1}  0  0  1 \bar{1}  0  0 \bar{1}  0  1 \bar{1}  0],\nonumber \\
&\cdots \mbox{their 16 partners} \nonumber \\
&65:[0 0 1 1 0 0 1 1 0 0 0 0 0 0 0 0 0 0 1 1 0 0 1 1 0 0 0 0 0 0 0 0],
66:[1 1 0 0 1 1 0 0 0 0 0 0 0 0 0 0 1 1 0 0 1 1 0 0 0 0 0 0 0 0 0 0],\nonumber \\
&67:[ 0  0  0  0  0  0  0  0  1 \bar{1}  0  0  1 \bar{1}  0  0  0  0  0  0  0  0  0  0  1 \bar{1}  0  0  1 \bar{1}  0  0],
68:[ 0  0  1 \bar{1}  0  0  1 \bar{1}  0  0  0  0  0  0  0  0  0  0 \bar{1}  1  0  0 \bar{1}  1  0  0  0  0  0  0  0  0],\nonumber \\
&69:[ 0  0  0  0  0  0  0  0  0  0  1  1  0  0 \bar{1} \bar{1}  0  0  0  0  0  0  0  0  0  0  1  1  0  0 \bar{1} \bar{1}],
70:[ 0  0  1  1  0  0 \bar{1} \bar{1}  0  0  0  0  0  0  0  0  0  0  1  1  0  0 \bar{1} \bar{1}  0  0  0  0  0  0  0  0],\nonumber \\
&71:[ 0  0  0  0  0  0  0  0  1  1  0  0 \bar{1} \bar{1}  0  0  0  0  0  0  0  0  0  0 \bar{1} \bar{1}  0  0  1  1  0  0],
72:[ 1  1  0  0 \bar{1} \bar{1}  0  0  0  0  0  0  0  0  0  0 \bar{1} \bar{1}  0  0  1  1  0  0  0  0  0  0  0  0  0  0],\nonumber \\
&73:[ 0  0  0  0  0  0  0  0  0  0  1 \bar{1}  0  0  1 \bar{1}  0  0  0  0  0  0  0  0  0  0 \bar{1}  1  0  0 \bar{1}  1],
74:[ 0  0  0  0  0  0  0  0  1  1  0  0  1  1  0  0  0  0  0  0  0  0  0  0 \bar{1} \bar{1}  0  0 \bar{1} \bar{1}  0  0],\nonumber \\
&75:[ 1  1  0  0  1  1  0  0  0  0  0  0  0  0  0  0 \bar{1} \bar{1}  0  0 \bar{1} \bar{1}  0  0  0  0  0  0  0  0  0  0],
76:[ 0  0  0  0  0  0  0  0  1 \bar{1}  0  0 \bar{1}  1  0  0  0  0  0  0  0  0  0  0 \bar{1}  1  0  0  1 \bar{1}  0  0],\nonumber \\
&77:[ 0  0  0  0  0  0  0  0  1 \bar{1}  0  0 \bar{1}  1  0  0  0  0  0  0  0  0  0  0  1 \bar{1}  0  0 \bar{1}  1  0  0],
78:[ 1 \bar{1}  0  0  1 \bar{1}  0  0  0  0  0  0  0  0  0  0  1 \bar{1}  0  0  1 \bar{1}  0  0  0  0  0  0  0  0  0  0],\nonumber \\
&79:[ 1 \bar{1}  0  0 \bar{1}  1  0  0  0  0  0  0  0  0  0  0  1 \bar{1}  0  0 \bar{1}  1  0  0  0  0  0  0  0  0  0  0],
80:[ 0  0  0  0  0  0  0  0  0  0  1 \bar{1}  0  0 \bar{1}  1  0  0  0  0  0  0  0  0  0  0 \bar{1}  1  0  0  1 \bar{1}],\nonumber \\
&\cdots \mbox{their 16 partners} \nonumber \\
&97:[1 0 1 0 0 0 0 0 1 0 1 0 0 0 0 0 1 0 1 0 0 0 0 0 1 0 1 0 0 0 0 0],
98:[0 0 0 0 1 0 1 0 0 0 0 0 1 0 1 0 0 0 0 0 1 0 1 0 0 0 0 0 1 0 1 0],\nonumber \\
&99:[ 1  0  1  0  0  0  0  0 \bar{1}  0 \bar{1}  0  0  0  0  0 \bar{1}  0 \bar{1}  0  0  0  0  0  1  0  1  0  0  0  0  0],
100:[ 0  0  0  0  1  0 \bar{1}  0  0  0  0  0 \bar{1}  0  1  0  0  0  0  0 \bar{1}  0  1  0  0  0  0  0  1  0 \bar{1}  0],\nonumber \\
&101:[ 0  0  0  0  1  0 \bar{1}  0  0  0  0  0  1  0 \bar{1}  0  0  0  0  0 \bar{1}  0  1  0  0  0  0  0 \bar{1}  0  1  0],
102:[ 1  0 \bar{1}  0  0  0  0  0  1  0 \bar{1}  0  0  0  0  0  1  0 \bar{1}  0  0  0  0  0  1  0 \bar{1}  0  0  0  0  0],\nonumber \\
&103:[ 0  0  0  0  0  1  0  1  0  0  0  0  0 \bar{1}  0 \bar{1}  0  0  0  0  0  1  0  1  0  0  0  0  0 \bar{1}  0 \bar{1}],
104:[ 0  0  0  0  1  0  1  0  0  0  0  0  1  0  1  0  0  0  0  0 \bar{1}  0 \bar{1}  0  0  0  0  0 \bar{1}  0 \bar{1}  0],\nonumber \\
&105:[ 0  0  0  0  0  1  0 \bar{1}  0  0  0  0  0 \bar{1}  0  1  0  0  0  0  0 \bar{1}  0  1  0  0  0  0  0  1  0 \bar{1}],
106:[ 0  0  0  0  0  1  0  1  0  0  0  0  0  1  0  1  0  0  0  0  0 \bar{1}  0 \bar{1}  0  0  0  0  0 \bar{1}  0 \bar{1}],\nonumber \\
&107:[ 1  0 \bar{1}  0  0  0  0  0  1  0 \bar{1}  0  0  0  0  0 \bar{1}  0  1  0  0  0  0  0 \bar{1}  0  1  0  0  0  0  0],
108:[ 0  0  0  0  1  0 \bar{1}  0  0  0  0  0  1  0 \bar{1}  0  0  0  0  0  1  0 \bar{1}  0  0  0  0  0  1  0 \bar{1}  0],\nonumber \\
&109:[ 0  0  0  0  1  0  1  0  0  0  0  0 \bar{1}  0 \bar{1}  0  0  0  0  0 \bar{1}  0 \bar{1}  0  0  0  0  0  1  0  1  0],
110:[ 0  0  0  0  0  1  0 \bar{1}  0  0  0  0  0 \bar{1}  0  1  0  0  0  0  0  1  0 \bar{1}  0  0  0  0  0 \bar{1}  0  1],\nonumber \\
&111:[ 0  1  0 \bar{1}  0  0  0  0  0 \bar{1}  0  1  0  0  0  0  0  1  0 \bar{1}  0  0  0  0  0 \bar{1}  0  1  0  0  0  0],
112:[ 0  1  0  1  0  0  0  0  0 \bar{1}  0 \bar{1}  0  0  0  0  0  1  0  1  0  0  0  0  0 \bar{1}  0 \bar{1}  0  0  0  0],\nonumber \\
&\cdots \mbox{their 16 partners} \nonumber \\
&129:[0 0 0 0 0 0 0 0 0 0 0 0 0 0 0 0 1 1 0 0 1 1 0 0 1 1 0 0 1 1 0 0],
130:[1 1 0 0 1 1 0 0 1 1 0 0 1 1 0 0 0 0 0 0 0 0 0 0 0 0 0 0 0 0 0 0],\nonumber \\
&131:[ 0  0  0  0  0  0  0  0  0  0  0  0  0  0  0  0  1  1  0  0  1  1  0  0 \bar{1} \bar{1}  0  0 \bar{1} \bar{1}  0  0],
132:[ 1 \bar{1}  0  0 \bar{1}  1  0  0 \bar{1}  1  0  0  1 \bar{1}  0  0  0  0  0  0  0  0  0  0  0  0  0  0  0  0  0  0],\nonumber \\
&133:[ 1  1  0  0  1  1  0  0 \bar{1} \bar{1}  0  0 \bar{1} \bar{1}  0  0  0  0  0  0  0  0  0  0  0  0  0  0  0  0  0  0],
134:[ 0  0  0  0  0  0  0  0  0  0  0  0  0  0  0  0  0  0  1 \bar{1}  0  0  1 \bar{1}  0  0 \bar{1}  1  0  0 \bar{1}  1],\nonumber \\
&135:[ 0  0  0  0  0  0  0  0  0  0  0  0  0  0  0  0  0  0  1 \bar{1}  0  0 \bar{1}  1  0  0  1 \bar{1}  0  0 \bar{1}  1],
136:[ 0  0  1  1  0  0 \bar{1} \bar{1}  0  0  1  1  0  0 \bar{1} \bar{1}  0  0  0  0  0  0  0  0  0  0  0  0  0  0  0  0],\nonumber \\
&137:[ 0  0  0  0  0  0  0  0  0  0  0  0  0  0  0  0  1 \bar{1}  0  0  1 \bar{1}  0  0  1 \bar{1}  0  0  1 \bar{1}  0  0],
138:[ 0  0  0  0  0  0  0  0  0  0  0  0  0  0  0  0  1  1  0  0 \bar{1} \bar{1}  0  0 \bar{1} \bar{1}  0  0  1  1  0  0],\nonumber \\
&139:[ 0  0  0  0  0  0  0  0  0  0  0  0  0  0  0  0  0  0  1  1  0  0 \bar{1} \bar{1}  0  0  1  1  0  0 \bar{1} \bar{1}],
140:[ 0  0  0  0  0  0  0  0  0  0  0  0  0  0  0  0  0  0  1  1  0  0 \bar{1} \bar{1}  0  0 \bar{1} \bar{1}  0  0  1  1],\nonumber \\
&141:[ 0  0  0  0  0  0  0  0  0  0  0  0  0  0  0  0  1 \bar{1}  0  0 \bar{1}  1  0  0  1 \bar{1}  0  0 \bar{1}  1  0  0],
142:[ 0  0  1 \bar{1}  0  0 \bar{1}  1  0  0 \bar{1}  1  0  0  1 \bar{1}  0  0  0  0  0  0  0  0  0  0  0  0  0  0  0  0],\nonumber \\
&143:[ 0  0  0  0  0  0  0  0  0  0  0  0  0  0  0  0  0  0  1 \bar{1}  0  0  1 \bar{1}  0  0  1 \bar{1}  0  0  1 \bar{1}],
144:[ 0  0  0  0  0  0  0  0  0  0  0  0  0  0  0  0  1 \bar{1}  0  0  1 \bar{1}  0  0 \bar{1}  1  0  0 \bar{1}  1  0  0],\nonumber \\
&\cdots \mbox{their 16 partners}. \nonumber \\
\end{eqnarray*}
\normalsize
For the sake of completeness, we also give a list of the $21$ maximal vector bases:
\small
\begin{eqnarray*}
&  1: \{ 65, 66, 67, 71, 72, 74, 75, 78, 80, 81, 82, 84, 85, 86, 89, 90,\nonumber \\
&~~~~~ 91, 92, 93, 95, 132, 134, 136, 139, 140, 141, 143, 151, 153, 155, 158, 160 \},\nonumber \\
&  2: \{ 67, 69, 70, 71, 72, 76, 77, 78, 79, 80, 84, 85, 86, 87, 88, 89,\nonumber \\
&~~~~~  92, 93, 95, 96, 129, 130, 131, 133, 134, 143, 145, 146, 147, 149, 153, 160 \},\nonumber \\
&  3: \{ 36, 37, 40, 43, 45, 46, 49, 50, 51, 54, 55, 57, 58, 60, 63, 64,\nonumber \\
&~~~~~  80, 92, 93, 95, 137, 138, 144, 145, 146, 147, 149, 150, 152, 154, 156, 159 \},\nonumber \\
&  4: \{ 2, 4, 5, 7, 10, 12, 13, 15, 18, 20, 21, 23, 26, 28, 29, 31, \nonumber \\
&~~~~~ 97, 99, 102, 103, 105, 106, 107, 110, 113, 115, 118, 119, 121, 122, 123, 126 \},\nonumber \\
&  5: \{ 33, 34, 38, 44, 47, 53, 56, 62, 66, 67, 68, 71, 73, 74, 75, 80,\nonumber \\
&~~~~~ 81, 82, 83, 85, 89, 90, 93,94, 132, 136, 139, 140, 141, 151, 155, 158 \},\nonumber \\
&  6: \{ 66, 68, 69, 70, 71, 72, 73, 74, 75, 80, 81, 83, 85, 86, 87, 88,\nonumber \\
&~~~~~  92, 93, 94, 95, 132, 137, 141, 144, 145, 146, 147, 149, 150, 151, 158, 159 \},\nonumber \\
&  7: \{ 65, 66, 67, 69, 70, 77, 78, 79, 81, 82, 83, 85, 86, 93, 94, 95,\nonumber \\
&~~~~~ 99, 100, 101, 104, 105, 106, 107, 109, 115, 116, 117, 120, 121, 122, 123, 125 \},\nonumber \\
&  8: \{ 36, 37, 43, 45, 50, 54, 57, 63, 67, 78, 84, 89, 129, 133, 134, 135,\nonumber \\
&~~~~~  138, 139, 142, 145, 146, 147, 148, 149, 151, 152, 153, 154, 155, 156, 157, 158 \},\nonumber \\
&  9: \{ 34, 38, 41, 47, 52, 53, 59, 61, 68, 69, 70, 73, 76, 77, 79, 83, \nonumber \\
&~~~~~ 87, 88, 94, 96, 129, 130, 131, 133, 135, 137, 138, 142, 145, 149, 150, 154 \},\nonumber \\
&  10: \{ 2, 3, 5, 8, 9, 12, 14, 15, 17, 20, 22, 23, 26, 27, 29, 32, \nonumber \\
&~~~~~ 33, 34, 37, 40, 41, 42, 43, 45, 47, 48, 51, 52, 54, 55, 60, 62 \},\nonumber \\
&  11: \{ 36, 43, 45, 51, 55, 57, 58, 64, 65, 67, 68, 70, 71, 72, 75, 76,\nonumber \\
&~~~~~  78, 79, 80, 81, 82, 84, 85, 86, 88, 89, 90, 91, 92, 93, 94, 95 \},\nonumber \\
&  12: \{ 66, 74, 75, 80, 81, 92, 93, 95, 132, 134, 136, 137, 138, 139, 140, 141,\nonumber \\
&~~~~~  143, 144, 145, 146, 147, 149, 150, 151, 152, 153, 154, 155, 156, 158, 159, 160 \},\nonumber \\
&  13: \{ 19, 20, 23, 24, 27, 28, 31, 32, 66, 67, 74, 75, 76, 77, 78, 79,\nonumber \\
&~~~~~  81, 84, 89, 96, 136, 138, 142, 145, 147, 152, 153, 154, 155, 156, 157, 160 \},\nonumber \\
&  14:  \{ 1, 2, 3, 4, 9, 10, 11, 12, 17, 18, 19, 20, 25, 26, 27, 28, \nonumber \\
&~~~~~ 98, 100, 101, 103, 104, 105, 106,08, 109, 110, 113, 115, 118, 123, 127, 128 \},\nonumber \\
&  15:  \{ 1, 2, 5, 6, 17, 18, 21, 22, 65, 67, 69, 70, 71, 74, 76, 77,\nonumber \\
&~~~~~  81, 82, 84, 86, 87, 88, 90, 91,  134, 135, 142, 143, 148, 153, 157, 160 \},\nonumber \\
&  16:  \{ 99, 100, 103, 105, 109, 110, 111, 112, 115, 116, 119, 121, 125, 126, 127, 128, \nonumber \\
&~~~~~ 129, 130, 135, 136, 137, 139, 141, 143, 145, 146, 151, 152, 153, 154, 157, 159 \},\nonumber \\
&  17:  \{ 11, 12, 15, 16, 17, 18, 21, 22, 25, 26, 27, 28, 29, 30, 31, 32,\nonumber \\
&~~~~~  65, 68, 70, 83, 87, 90, 92, 93, 130, 132, 133, 150, 151, 154, 156, 159 \},\nonumber \\
&  18:  \{ 33, 35, 36, 37, 38, 39, 41, 46, 49, 51, 52, 53, 54, 55, 57, 62,\nonumber \\
&~~~~~  98, 100, 103, 106, 107, 108,  109, 110, 114, 116, 119, 122, 123, 124, 125, 126 \},\nonumber \\
&  19: \{ 67, 69, 70, 71, 72, 78, 80, 84, 85, 86, 87, 88, 89, 92, 93, 95,\nonumber \\
&~~~~~  129, 130, 131, 132, 133, 134, 141, 143, 145, 146, 147, 149, 151, 153, 158, 160 \},\nonumber \\
&  20:  \{ 33, 35, 36, 37, 39, 40, 42, 43, 44, 45, 46, 48, 49, 50, 51, 54,\nonumber \\
&~~~~~  55, 56, 57, 58, 60, 62, 63,  64, 144, 146, 147, 148, 152, 156, 157, 159 \},\nonumber \\
&  21:  \{ 1, 3, 6, 8, 9, 11, 14, 16, 17, 19, 22, 24, 25, 27, 30, 32,\nonumber \\
&~~~~~  98, 100, 101, 104, 108, 109, 111, 112, 114, 116, 117, 120, 124, 125, 127, 128 \}. \nonumber \\
\end{eqnarray*}
\normalsize
\noindent
Rays $1$ to $32$ pertain to the second (computational) basis in (\ref{BKSstate}), the subsequent rays correspond to the remaining four bases. Each  of the five aggregates of rays
contains a partner aggregate (not made explicit), which means that our BKS state proof is also Bell's proof.
To arrive at the $160-21$ proof, we randomly selected a small set $S$ of bases among the $661$ ones such that (a) there is at least one subset of $S$ containing $5$ bases partitioning the $5 \times 32=160$ rays (this criterion was simply adopted to reach the desired result with only $32^5$ checks), (b) the set $S$ itself satisfies the BKS `non-coloring' constraints given in the introduction. Applying this methodology in a recursive way, we arrived at the $21$ maximal bases which contain a single subset $\{1,4,9,20,21\}$ partitioning the rays.

\subsection{Five-qubit contextuality and a distribution of distances between maximal bases}

\begin{figure}[t]
\centering
\includegraphics[width=11cm,clip=]{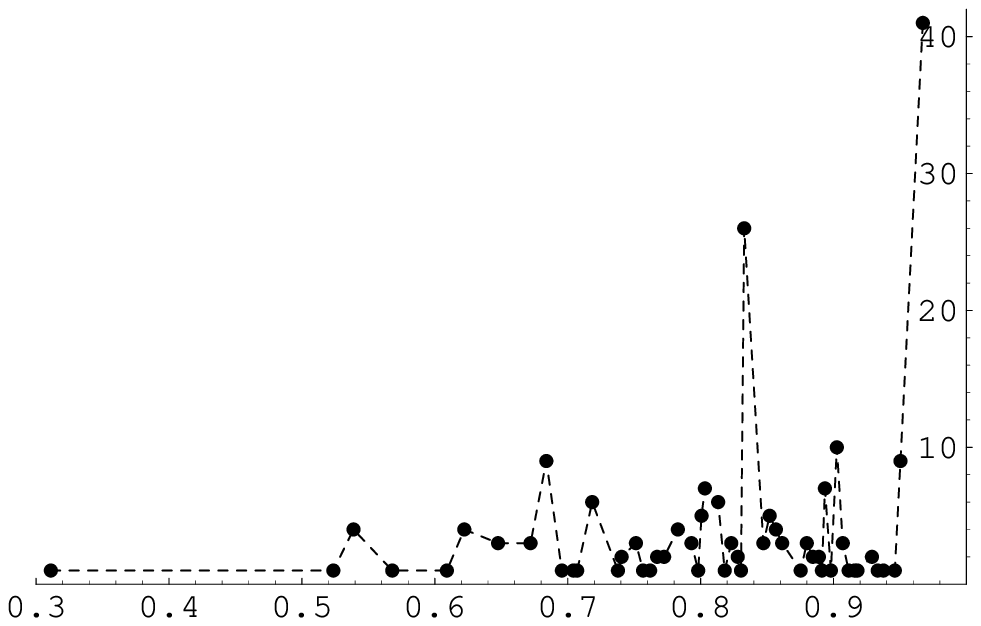}\\ \includegraphics[width=11cm,clip=]{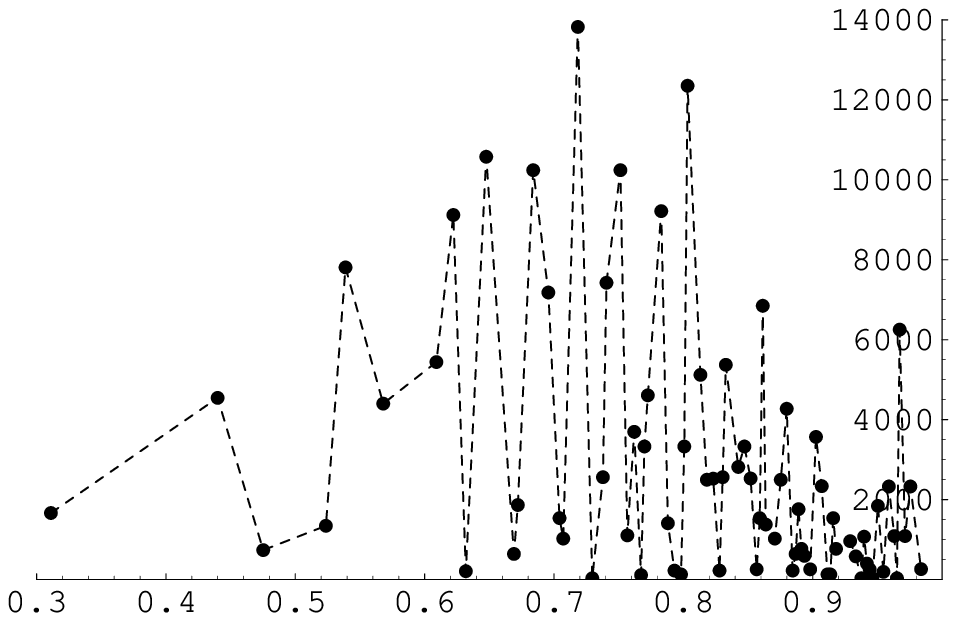}
\caption{{\it Top:} -- A histogram of distances between the maximal bases for the $160-21$ proof of the BKS theorem. There are two main peaks located at  $d_1=(29/31)^{1/2}\sim 0.9672$ and $d_2=(43/62)^{1/2}\sim 0.8328$, but the distances are spread over $54$ distinct values exhibiting a noise-like distribution; this is quite remarkable given the fact that the graph described in Sec. 2.1 possesses a rather high degree of symmetry, $\mathbb{Z}_2^5 \rtimes \mathbb{Z}_6$. {\it Bottom:} -- A histogram of distances between the maximal bases for the $160-661$ proof. Here the distances acquire as many as $77$ distinct values, whose distribution looks again like a noise. Comparing with the top figure, one observes
that a bunch of peaks in the middle range of distances is redundant for the present BKS proof because the characteristic peaks at $d_1$ and $d_2$ are already very well discernible.}
\label{fig1}
\end{figure}

Apart from the use of standard graph theoretical tools for characterizing the ray/base symmetries, we can also analyze our sets in terms of the `Hilbert-Schmidt'
distance $D_{ab}$ between two orthonormal bases $a$ and $b$, defined as \cite[eq. (2)]{Raynal2011}-\cite{Beng2007}
\begin{equation*}
D_{ab}^2=1-\frac{1}{d-1}\sum_{i,j=1}^d \left(\left|\left\langle a_i|b_j \right\rangle\right|^2-\frac{1}{d}\right)^2.
\label{Beng}
\end{equation*}
This distance vanishes when the bases are the same and is maximal (and equal to unity) when the two bases $a$ and $b$ are mutually unbiased, $\left|\left\langle a_i|b_j \right\rangle\right|^2=1/d$, and only then. It has already been found \cite{Planat2011} that the bases yielding a BKS proof prefer a particular set/pattern of distances, which we suspect to be a universal feature of such a proof. For the present proof(s), we again observe a good wealth of distances between the maximal bases that exhibit a remarkable noise-like pattern, as illustrated in Fig. \ref{fig1}.

\begin{figure}[t]
\centerline{\includegraphics[width=5cm,clip=]{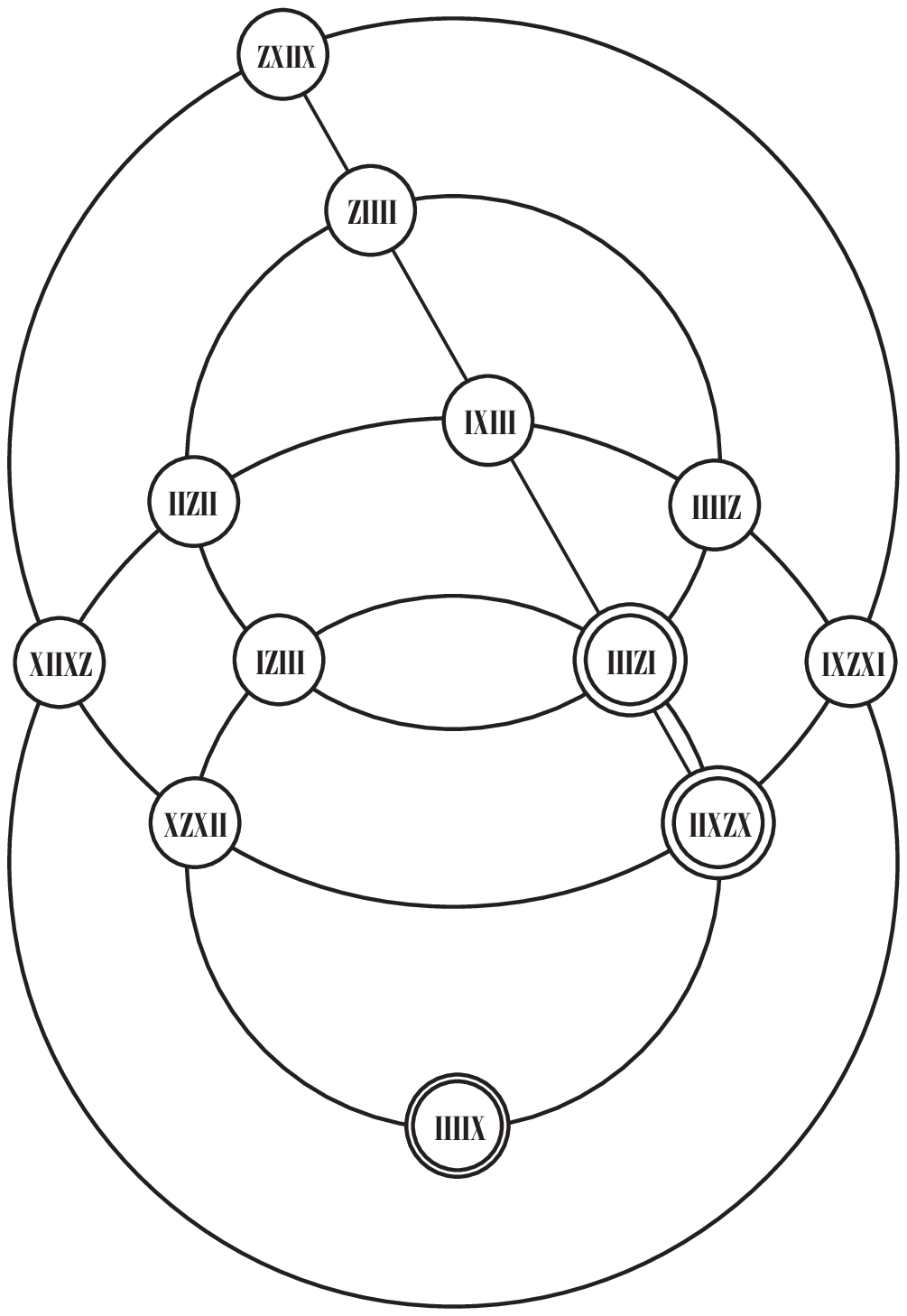}~~~~~~~\includegraphics[width=5cm,clip=]{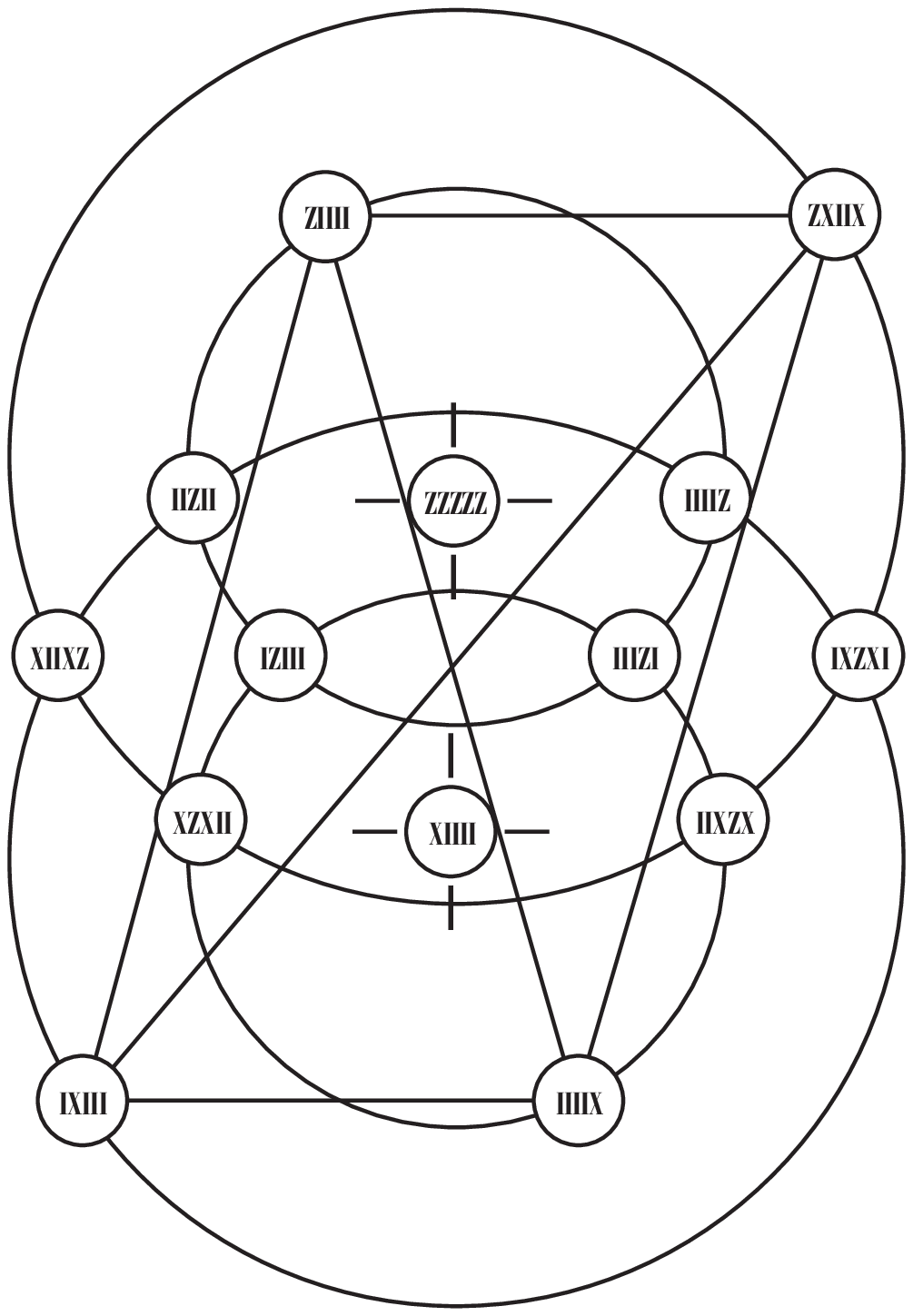}}
\vspace*{.2cm}
\caption{{\it Left:} -- An illustration of the relations between the five sets of five-qubit observables of (\ref{BKSstate}). Two big concentric circles at the top represent sets $A$ and $B$, those at the bottom sets $A'$ and $B'$; set $C$ is represented by the line-segment. The three ``exceptional/distinguished" observables are indicated by small double-circles. {\it Right:} -- The same for the five sets of (\ref{BKSoperator}); note that this latter configuration is more symmetric that the former one.
In this section, we use a full-fledged labeling of the observables and a shorthand notation for the tensor product, e.\,g. $XZXII \equiv X \otimes Z \otimes X \otimes I \otimes I$.}
\end{figure}
\subsection{Five-qubit contextuality and finite geometry}
In this section we shall provide the reader with a  finite-geometrical insight into the structure of the five sets (\ref{BKSstate}). To this end in view, it is instructive to represent mutual relations between these sets in a diagrammatical form as depicted in Fig.\,2, {\it left}. One first notes that the observables $IIIZI$ and $IIXZX$ belong to three different sets, whilst the observable $IIIIX$ sits in just one set; all the remaining elements are in exactly two sets. We shall, however, be more interested in geometry of each set as a whole. In this respect it is fairly obvious that we have two pairs of sets, $A - B$ and $A' - B'$, and that set $C$ stands on a different footing with respect to each of the two pairs.

To find a finite-geometrical underpinning of this relation, we shall invoke some of the theory of factor-group-generated finite polar spaces expounded thoroughly in \cite{hos}, where the interested reader is referred to look for all the necessary background information and more details (see also \cite{sp07,ps08,thas}). All our observables represent elements from the real five-qubit Pauli group, whose geometry is that of the symplectic polar space $W(9,2)$. This space, roughly speaking, is a collection of all totally isotropic subspaces of the ambient nine-dimensional binary projective space, PG$(9,2)$, equipped with a non-degenerate alternating bilinear form.
The elements of the group are in a bijective correspondence with the points of $W(9,2)$ in such a way that two commuting elements correspond to two points joined by a totally isotropic line; a maximum set of mutually commuting elements of the group having its counterpart in a maximal totally isotropic subspace (also called a generator), which is PG$(4,2)$.

\begin{table}[t]
\small
\begin{center}
\caption{The points/observables of the  five PG$(4,2)$s. Here, for example, `123' stands for the product of observables `1', `2' and `3'. } \vspace*{0.2cm}
\begin{tabular}{rccccc}
\hline \hline
Point   & [$A$]  & [$B$] & [$A'$]  & [$B'$] & [$C$]\\
\hline
1      & $XZXII$  & $IZIII$ & $IZIII$  & $IIZII$ & $IXIII$\\
2      & $IXZXI$  & $IIZII$ & $IIIZI$  & $IIIIZ$ & $IIIZI$\\
3      & $IIXZX$  & $IIIZI$ & $XZXII$  & $XIIXZ$ & $ZIIII$\\
4      & $XIIXZ$  & $IIIIZ$ & $IIXZX$  & $IXZXI$ & $ZXIIX$\\
5      & $ZXIIX$  & $ZIIII$ & $IIIIX$  & $IXIII$ & $IIXZX$\\
12     & $XYYXI$  & $IZZII$ & $IZIZI$  & $IIZIZ$ & $IXIZI$\\
13     & $XZIZX$  & $IZIZI$ & $XIXII$  & $XIZXZ$ & $ZXIII$\\
14     & $IZXXZ$  & $IZIIZ$ & $IZXZX$  & $IXIXI$ & $ZIIIX$\\
15     & $YYXIX$  & $ZZIII$ & $IZIIX$  & $IXZII$ & $IXXZX$\\
23     & $IXYYX$  & $IIZZI$ & $XZXZI$  & $XIIXI$ & $ZIIZI$\\
24     & $XXZIZ$  & $IIZIZ$ & $IIXIX$  & $IXZXZ$ & $ZXIZX$\\
25     & $ZIZXX$  & $ZIZII$ & $IIIZX$  & $IXIIZ$ & $IIXIX$\\
34     & $XIXYY$  & $IIIZZ$ & $XZIZX$  & $XXZIZ$ & $IXIIX$\\
35     & $ZXXZI$  & $ZIIZI$ & $XZXIX$  & $XXIXZ$ & $ZIXZX$\\
45     & $YXIXY$  & $ZIIIZ$ & $IIXZI$  & $IIZXI$ & $ZXXZI$\\
123    & $XYZYX$  & $IZZZI$ & $XIXZI$  & $XIZXI$ & $ZXIZI$\\
124    & $IYYIZ$  & $IZZIZ$ & $IZXIX$  & $IXIXZ$ & $ZIIZX$\\
125    & $YZYXX$  & $ZZZII$ & $IZIZX$  & $IXZIZ$ & $IXXIX$\\
134    & $IZIYY$  & $IZIZZ$ & $XIIZX$  & $XXIIZ$ & $IIIIX$\\
135    & $YYIZI$  & $ZZIZI$ & $XIXIX$  & $XXZXZ$ & $ZXXZX$\\
145    & $ZYXXY$  & $ZZIIZ$ & $IZXZI$  & $IIIXI$ & $ZIXZI$\\
234    & $XXYZY$  & $IIZZZ$ & $XZIIX$  & $XXZII$ & $IXIZX$\\
235    & $ZIYYI$  & $ZIZZI$ & $XZXZX$  & $XXIXI$ & $ZIXIX$\\
245    & $YIZIY$  & $ZIZIZ$ & $IIXII$  & $IIZXZ$ & $ZXXII$\\
345    & $YXXYZ$  & $ZIIZZ$ & $XZIZI$  & $XIZIZ$ & $IXXZI$\\
1234   & $IYZZY$  & $IZZZZ$ & $XIIIX$  & $XXIII$ & $IIIZX$\\
1235   & $YZZYI$  & $ZZZZI$ & $XIXZX$  & $XXZXI$ & $ZXXIX$\\
1245   & $ZZYIY$  & $ZZZIZ$ & $IZXII$  & $IIIXZ$ & $ZIXII$\\
1345   & $ZYIYZ$  & $ZZIZZ$ & $XIIZI$  & $XIIIZ$ & $IIXZI$\\
2345   & $YIYZZ$  & $ZIZZZ$ & $XZIII$  & $XIZII$ & $IXXII$\\
12345  & $ZZZZZ$  & $ZZZZZ$ & $XIIII$  & $XIIII$ & $IIXII$\\
\hline \hline
\end{tabular}
\end{center}
\end{table}

Next, a PG(4,2) has 31 points (see, for example, \cite{hirsch}). If we multiply the elements and their products within each set of (\ref{BKSstate}), we also get 31 distinct values, which means that each of our five sets spans a PG$(4,2)$ in $W(9,2)$. Table 1 lists explicitly the set of 31 observables/points for each of these  PG$(4,2)$s; here, $[A]$ is a shorthand for the PG$(4,2)$ spanned by $A$, etc.
Furthermore, as all elements in each of the five PG$(4,2)$s are symmetric, these spaces at the same time correspond to generators on the hyperbolic quadric $Q^{+}(9,2)$ that is the locus of  symmetric elements of the  group \cite[$\S$ 8]{hos}. As it is well known \cite[$\S$ 22.4]{hith}, such a quadric features two systems of generators, with two different generators pertaining to the same system if they share a projective space of dimension 2 (plane) or 0 (point), and to different systems if this dimension is 3 (solid), 1 (line) or $-1$ (an empty set). Employing Table 1, one finds that our five PG$(4,2)$s have the following intersection properties:\\

$[A] \cap [B] = \{ZZZZZ\}$,

$[A] \cap [A'] = \{XZIZX, XZXII, IIXZX\}$,

$[A] \cap [B'] = \{XXZIZ, XIIXZ, IXZXI\}$,

$[A] \cap [C] = \{ZXXZI, ZXIIX, IIXZX\}$,

$[B] \cap [A'] = \{IZIZI, IZIII, IIIZI\}$,

$[B] \cap [B'] = \{IIZIZ, IIZII, IIIIZ\}$,

$[B] \cap [C] = \{ZIIZI, IIIZI, ZIIII\}$,

$[A'] \cap [B'] = \{XIIII\}$,

$[A'] \cap [C] = \{IIXIX, IIIZX, IIXZI, IIXII, IIIZI, IIXZX, IIIIX\}$,

$[B'] \cap [C] = \{IXIII\}$.\\
\\
Here, each three-element set represents a line and the seven-element one represents a plane. Rephrased in the language of dimensions, the relations between the five spanned PG$(4,2)$s read
as shown in Table 2.
\begin{table}[t]
\begin{center}
\caption{The dimensions of projective spaces of pairwise intersections of the  five PG$(4,2)$s.} \vspace*{0.2cm}
\begin{tabular}{l|ccccc}
\hline \hline
   & $[A]$  & $[B]$ & $[A']$  & $[B']$ & $[C]$\\
\hline
$[A]$      & --  & 0 & 1  & 1 & 1\\
$[B]$      & 0  & -- & 1  & 1 & 1\\
$[A']$     & 1  & 1 & --  & 0 & 2\\
$[B']$     & 1  & 1 & 0  & -- & 0\\
$[C]$      & 1  & 1 & 2  & 0 & --\\
\hline \hline
\end{tabular}
\end{center}
\end{table}
We see that $A$-space and $B$-space are in the same system, as are $A'$- and $B'$-spaces, the two systems being different; this accounts for the pairing property mentioned above (see Fig. 2).
We further see that although $C$-space lies in the same system as $A'$- and $B'$-ones, it has different intersection with each of the latter; this explains why set $C$ has a different footing as well. One also observes that the three distinguished observables (see Fig. 2) are all accommodated by the unique Fano plane $[A'] \cap [C]$.

\section{Conclusion}
We proposed particular $160-21$ and $160-661$ five-qubit state proofs of the BKS theorem, each of which also furnishes a proof of Bell's theorem, and studied their essential features. The `Hilbert-Schmidt' distances between the corresponding maximal bases show a noise-like distribution; this is quite remarkable especially in the second case, where the graph whose vertices are the 21 bases and edges join two bases whenever they overlap  exhibits a relatively high degree of symmetry, $\mathbb{Z}_2^5 \rtimes \mathbb{Z}_6$. We also came across a rather counter-intuitive feature that our starting ``magic" configuration of observables (eq.\,(1)) does not yield a state proof, and we had to pass to a slightly different one (eq.\,(2)) to do the job.  The geometric nature of the latter latter configuration was clarified in terms of symplectic geometry $W(9,2)$ and its refinement, the hyperbolic quadric  $Q^{+}(9,2)$ that is the locus of  symmetric elements of the real five-qubit Pauli group. We expect this approach to state proofs of the BKS theorem, which combines group-theoretical tools with finite-geometrical reasoning, to be very promising especially for $N$-qubits with growing values of $N$, where we surmise the noise-like behavior to be more pronounced and the corresponding finite-geometric underpinning more complex/intricate.

\section*{Acknowledgements}
This work was partially supported by the VEGA grant agency project 2/0098/10. We are extremely grateful to our friend Petr Pracna for an electronic version of Figure 2.

\end{document}